\documentclass[prd,twocolumn,showpacs,showkeys]{revtex4}
\usepackage[dvips]{graphicx}
\usepackage{dcolumn}
\usepackage{bm}

\usepackage{amssymb, latexsym, amsopn}

\begin{document}

\title{Evolution of cosmic structures in different environments in the
quasispherical Szekeres model}

\author{Krzysztof  Bolejko}
\affiliation{Nicolaus Copernicus Astronomical Center, Polish Academy of Sciences,
 Bartycka 18, 00-716 Warsaw, Poland}
\email{bolejko@camk.edu.pl}
\homepage{http://www.camk.edu.pl/~bolejko}

\date{\today}

\begin{abstract}
This paper investigates evolution of cosmic structures in different environments.
For this purpose the quasispherical Szekeres model is employed. The Szekeres model
is an exact solution of the Einstein field equations
within which it is possible to describe more than one structure. 
In this way investigations of the evolution of the cosmic structures presented 
here can be freed from such assumptions
as small value of the density contrast.
Also, studying the evolution of two or three structures within one framework
enables us to follow the interaction between these structures and their impact on
the evolution. Main findings include a conclusion that 
 small voids surrounded by large overdensities evolve slower than large, 
isolated voids do. On the other hand, large voids enhance the evolution 
of adjacent galaxy superclusters which evolve much faster than isolated superclusters.
\end{abstract}

\pacs{98.65.Dx, 98.65.-r, 04.20.Jb, 98.62.Ai}
\keywords{cosmology; structure formation; Szekeres model}

\maketitle

\section{Introduction}

At the end of 1970s astronomers provided observational evidence that
galaxies in the Universe are distributed inhomogeneously.
Galaxy redshift surveys show that the galaxies form structures 
such as voids, clusters and filaments.
Although this is only the distribution of visible matter,
there are some strong indications that
visible matter does trace the distribution of dark matter, so
 real matter distribution is similar.
 All these structures evolved from small initial fluctuations 
 which started to grow after the last scattering moment. 
 However, diffrent structures evolved in various ways,
depending on their environment and neighborhood.
 The present-day density contrast [Eq. (\ref{dencond})] of 
overdense regions is larger than 1
\cite{BE} and inside voids 
it decends to -1 \cite{Hoy}. Thus, these structures must be described by exact solutions of the Einstein equations without  such assumptions as small value of a density contrast.
This paper provides the analysis of cosmic structures'  evolution which 
is free of such assumptions.
The evolution of the cosmic structures in diffrent environments  
is investigated by employing the quasispherical Szekeres model which is
an exact solution of the  Einstein field equations.

The structure of this paper is as follows: Sec.
\ref{szekmdl} presents the Szekeres model;
Sec. \ref{doubstr} presents the evolution of 
pairs of voids and superclusters in the quasispherical Szekeres model;
Sec. \ref{expan} presents the role of expansion in
the process of structure formation; Sec. \ref{tripstr}
presents the connection between the results   obtained 
in the Szekeres model and the real large-scale structure of the Universe.

\section{The Szekeres model}\label{szekmdl}

For our purpose it is convenient to use a coordinate system 
which is different from
that in which Szekeres \cite{SZ1} originally found his solution. The metric is of
the following form \cite{HK}:

\begin{equation}
ds^2 =  c^2 dt^2 - \frac{(\Phi' - \Phi \frac{\textstyle E'}{\textstyle E})^2}
{(\varepsilon - k)} dr^2 - \Phi^2 \frac{(dp^2 + dq^2)}{E^2}, \label{ds2}
 \end{equation}
where ${}' \equiv \partial/\partial r$, $\Phi = \Phi(t,r)$,  $\varepsilon = \pm1,0$ and $k = k(r)
\leq \varepsilon$ is an arbitrary function of $r$.

The function $E$ is given by: 
 \begin{equation}
E(r,p,q) = \frac{1}{2S}(p^2 + q^2) - \frac{P}{S} p - \frac{Q}{S} q + C ,
 \end{equation}
where the functions $S = S(r)$, $P = P(r)$, $Q = Q(r)$, and $C = C(r)$ satisfy
the relation:
 \begin{equation}
C = \frac{P^2}{2S} + \frac{Q^2}{2S} + \frac{S}{2} \varepsilon,~~~~~~~~~
\varepsilon = 0, \pm 1,
 \end{equation}
but are otherwise arbitrary.

As can be seen from (\ref{ds2}), only $\varepsilon = +1$ allows the model to have all
three Friedmann limits (hyperbolic, flat, and  spherical). This 
is induced by the requirement of the Lorentzian signature of the metric (\ref{ds2}). As we are interested
in the Friedmann limit of our model, i.e. we expect it becomes a homogeneous
Friedmann model at a large distance from the origin, we will focus only on the
$\varepsilon = 1$ case. This case is 
 often called the quasispherical Szekeres model.

Applying metric (\ref{ds2}) to the Einstein equations, 
with the assumption that the energy momentum tensor describes dust,
 the Einstein equations reduce  to the following two:

\begin{equation}
\frac{1}{c^2} \dot{\Phi}^2 (t,r) = \frac{2M(r)}{\Phi(t,r)} - k(r) + \frac{1}{3} \Lambda
\Phi^2(t,r), \label{vel}
\end{equation}

\begin{eqnarray}
&& 4 \pi \frac{G}{c^2} \rho(t,r,p,q) \nonumber \\ 
&& = \frac{M'(r) - 3 M(r) E'(r,p,q)/E(r,p,q)}{\Phi^2(t,r) [ \Phi'(t,r) - \Phi(t,r) E'(r,p,q)/E(r,p,q)]}. \label{rho}
\end{eqnarray}

In a Newtonian limit $M c^2/G$ is equal to the mass inside the shell of radial coordinate
$r$.  However, it is not an integrated rest mass but rather active gravitational
 mass that generates a gravitational field.
Although the $\rho$ function in Eq. (\ref{rho})  is a function of all coordinates,
it can be shown that the density can be decomposed into two 
parts: the monopole distribution and the part which has
a dipole structure \cite{SZ2,dS,PK}:

\begin{equation}
\epsilon = \epsilon_{mon}(t,r) + \epsilon_{dip}(t,r,p,q).
\end{equation}

The function $k(r)$ is another arbitrary function defining the Szekeres model.
By analogy with the Newtonian energy conservation equation, Eq. (\ref{vel}) shows that the
function $(- k/2)$ represents the energy per unit mass of the particles in the
shells of matter at constant $r$. On the other hand, by analogy with the
Friedmann equation and from the metric (\ref{ds2}) the function $k$ determines the
geometry of the spatial sections $t = $ const. However, since $k$ is a function
of the radial coordinate the geometry of the space is now  position dependent.

Eq. (\ref{vel}) can be integrated:

\begin{equation}
\int\limits_0^{\Phi}\frac{{\rm d}
\tilde{\Phi}}{\sqrt{\frac{2M(r)}{\tilde{\Phi}} - k(r) + \frac{1}{3} \Lambda
\tilde{\Phi}^2}} = c \left[t- t_B(r)\right], \label{cal}
\end{equation}
where $t_B$ is an arbitrary function of
$r$. This means that the Big Bang is not a single event as in the Friedmann
models but occurs at different times for different distances from the origin.

As can be seen the Szekeres model is specified by 6 functions. However, by a
choice of the coordinates, the number of independent functions can be reduced
to 5.

The equations of motion $T^{\alpha \beta}{}_{; \beta} = 0$ are reduced to the
continuity equation:

\begin{equation}
\dot{\rho} + \rho \Theta = 0,
\label{coneq}
\end{equation}
where $\Theta$ is the scalar of expansion and is equal to

\begin{eqnarray}
&& \Theta(t,r,p,q)  = 3 \frac{\dot{\Phi}(t,r)}{\Phi(t,r)} \nonumber \\ 
&& + \frac{ \dot{\Phi}'(t,r) - 
\dot{\Phi}(t,r) \Phi'(t,r)/\Phi(t,r)}{\Phi'(t,r) - \Phi(t,r) E'(r,p,q)/E(r,p,q)} \label{eks}
\end{eqnarray}

In the expanding Universe $\Theta$ is positive
so the density decreases. The structures which  exist in the Universe,
emerged either due to slower expansion of the space (formation of 
overdense regions) or 
due to faster expansion (formation of underdense regions).
In the Friedmann limit $R \rightarrow r a$, where $a$ is the scale factor and $\Theta \rightarrow 3 H_0$.

The Szekeres model is known to have no symmetry \cite{BST}. It is of great flexibility and wide application in cosmology \cite{BT} 
and in astrophysics \cite{HK,SZ2}, and still it can be used as a model of many astronomical phenomena. 
In this paper it will be employed to study the evolution of cosmic structures in 
different environments.

\subsection{Density contrast}

To compare the evolution of different models the change in their
density contrast is going to be considered.
Two different types of density contrast indicators
are taken into account.

The first one is a usual density contrast defined as follows:

\begin{equation}
\delta = \frac{\rho - \rho_b}{\rho_b},
\label{dencond}
\end{equation}
where $\rho_b$ is the background density.

However, the density contrast defined as above is 
 a local quantity
and is not covariant with the coordinate transformations.
The spatially invariant  density contrast can be defined as follows 
\cite{MT}:

\begin{equation}
S_{IK} = \int_{\Sigma} \left| \frac{h^{\alpha \beta}}{\rho^I} \frac{\partial \rho}{ \partial x^{\alpha}}
\frac{\partial \rho}{ \partial x^{\beta}} \right|^K {\rm d} V,
\end{equation}
where $I \in \mathbb{R}$, and $K \in \mathbb{R} \backslash \{0\}$.
This family of the density contrast indicators can be
considered as local or global depending on the size of $\Sigma$.
Such a quantity not only describes the change of density
but also the change of gradients and the volume
of a perturbed region. So this density indicator 
describes the evolution of the whole region
in a more sophisticated way than the  $\delta$.
Here only the case $I=2, K=1/2$ will be considered.

 All models presented  in this paper 
are calculated numerically as, unfortunately, the class of models
described by an analytical solution is
not sufficiently enough 
to describe the considered cosmic structures.
However, in the $\varepsilon = 1$ (and in the $\varepsilon = 0$) case
the $p,q$ coordinates have an infinite range.
Therefore, it is more convenient to use a
diffrent coordinate system where coordinates 
do have a finite range.

\subsubsection{Coordinate system}

The surface of constat $t$ and $r$ can be
represented by a stereographic projection
Employing the stereographic projection we can map the infinite surface
of $p,q$ coordinates to a surface of a sphere which has a finite range of coordinates $\theta, \phi$.

After the following transformations,

 \begin{eqnarray}
 p - P &=& S {\rm cot} \left( \frac{ \theta }{2} \right) \cos (\phi), \nonumber
 \\
 q - Q &=& S {\rm cot} \left( \frac{ \theta }{2} \right) \sin (\phi), \nonumber
 \\
 r &=&  r,
 \label{sphtrsf}
 \end{eqnarray}

we obtain

\begin{equation}
\frac{1}{E^2} (dp^2 + dq^2) =  \left( {\rm d}{\theta}^2 + \sin^2
\theta {\rm d}{\phi}^2 \right).
\end{equation}

The metric (\ref{ds2}) after such transfromations becomes nondiagonal:

\begin{eqnarray}
&& {\rm d} s^2 = c {\rm d} t^2 - \left\{ \frac{( \Phi' - \Phi E'/E)^2}{1 -k}
+ \frac{\Phi^2}{E^2}  \left[ S'^2 \cot^2 \frac{\theta}{2}   
\right. \right. \nonumber \\
&&   +  2 S' \cot \frac{\theta}{2} \left( Q' \sin \phi + P' \cos \phi \right)  
\nonumber \\ 
&& \left. \left. +  \left(P'^2 + Q'^2 \right) \right] \right\} {\rm d} r^2 
-  \frac{\Phi^2}{E^2} \left[ 2 S \cot \frac{\theta}{2}  \left( Q'
\cos \phi \right. \right.  \nonumber \\
&&  \left. \left. - P' \sin \phi \right)  \right] {\rm d} r {\rm d} {\phi}
 +  2 \frac{\Phi^2}{E} \left(   Q' \sin \phi + P' \cos \phi 
\right. \nonumber \\
&& \left. + S' \cot \frac{\theta}{2} \right) {\rm d} r {\rm d} {\theta}  -  \Phi^2  \left( {\rm d}{\theta}^2 + \sin^2\theta {\rm d} {\phi}^2 \right),
\label{ds2ss}
\end{eqnarray}
where

\begin{equation}
\frac{1}{E} = \frac{1 - \cos \theta }{S},
\label{enu}
\end{equation}

and

\begin{equation}
\frac{E'}{E} = \frac{S' \cos \theta + \sin \theta \left(P' \cos \phi + Q'
\sin \phi \right)}{S}. \label{nuz}
\end{equation}

As can be seen, if $t$ = const and $r$ = const, the above  becomes the
metric of the 2--dimensional sphere. Hence,  every $t$ = const and $r$ = const
slice of the Szekeres $\varepsilon = 1$ space--time is a sphere. However, as
$S, P$ and $Q$ are now functions of $r$, the spheres are not concentric. For the
spheres to be concentric, the following conditions must hold:

\begin{eqnarray}
P' = 0, \nonumber \\
Q' = 0, \nonumber \\
S' = 0.
\label{sscon}
\end{eqnarray}
Such conditions entail spherical symmetry and the metric (\ref{ds2ss})
becomes the line element of the Lema\^itre--Tolman model \cite{Lem,Tol}.
Due to this non-concentricity of spheres the density distribution has a structure of a
time--dependent mass dipole superposed on a monopole. However, since $S, P$ and $Q$ are position dependent, the axis of the dipole also changes in the space. 
The functions $S, P$, and $Q$ describe the position of this dipole, and as can be seen from 
Eq. (\ref{nuz}), $S$ describes the vertical position of the dipole component, 
while $P$ and $Q$ describe its horizontal position.

\subsection{Model set--up}

To specify model 5 functions
of the radial coordinate
need to be known. Let us define the radial coordinate as a
value of $\Phi$ at the initial instant $t_0 =  0.5 \times 10^6$ yr  after the big bang:
\begin{equation}
\tilde{r}:= \Phi(r,t_0).
\end{equation}
However, for clarity in futher use,  
the $\tilde{}$ sign is omitted 
and the new radial
coordinate will be referred to as $r$.

Two of these functions will be $t_B(r)$ and $M(r)$.
Let us assume that $t_B(r) = 0$.
 The function $M(r)$ describes the active gravitational mass inside the $t = $
const, $r =$ const sphere. Let us describe the mass function in the following form:

\begin{equation}
M(r) = M_0(r) + \delta M(r),
\end{equation}
where $M_0$ is the mass distribution as in the homogeneous universe,
and $\delta M$ is a mass correction, which can be either positive or negative.
The $\delta M$ is defined similarly as in the spherical symmetric case:

\begin{equation}
\delta M(r) = 4 \pi \frac{G}{c^2} \int_0^r {\rm d} u \Phi^2(u,t_0) \Phi'(u,t_0) 
\delta \bar{\rho}(u),
\end{equation}
where $\delta \bar{\rho}(r)$ is an arbitrary function chosen to specify the $\delta M$.
Although $\delta \bar{\rho}(r)$ is not the initial function of density fluctuations (since an initial density fluctuation is a function of all coordinates) it gives some estimation on the initial density fluctuation of the 
monopole density component.

The next three functions are $P(r), Q(r), S(r)$.
 All functions defining the model are presented 
as each case is being considered.
The numerical algorithm used to solve the Szekeres model's equations
is presented in detail in Ref. \cite{kb2}.

The chosen background model is the homogeneous Friedmann model with the density:
 \begin{equation}
   \rho_b = \Omega_m \times \rho_{cr} = 0.24 \times \frac{3H_0^2}{8 \pi G}.
   \label{rbdf}
 \end{equation}
 where the Hubble constant is $H_0 =74$ km s$^{-1}$ Mpc$^{-1}$.  The cosmological constant,  $\Lambda$, corresponds to $\Omega_{\Lambda} = 0.76$, where $\Omega_{\Lambda} = (1/3)  ( c^2 \Lambda/H_0^2)$.

\section{Double structures}\label{doubstr}

In this section the evolution of double structures,
namely a void with an adjacent galaxy supercluster, is investigated.
Although within the Szekeres model 
more than two structures can be described,
such investigations of less complex cases may also be useful
because they enable us to draw some general conclusions
without going into too much detail which could easily
obscure the larger picture.
Even then in Sec. \ref{tripstr} a more complex model is also
investigated and it is found that the rules extracted on the basis of the investigations of the double structures are still valid for such more complex situations.
The evolution of a double structure was also previously 
investigated by Bolejko \cite{kb3}.
However, the analysis presented in this paper 
is much more detailed and comprehensive.

\subsection{Models with $P' = 0 = S',~Q' \ne 0$}\label{p's'z}

As mentioned above, if $P' = 0 = S' = Q'$ the Szekeres model becomes
the Lema\^itre--Tolman model. Hence, the class of models considered 
in this subsection is the simplest generalisation of 
the spherically symmetric models.

The double structure of a void and adjacent supercluster 
can be described in the Szekeres model in two different ways.
The first alternative is when $\delta M<0$, the second when $\delta M>0$.
Both these possibilities are examined here.

\subsubsection{Models specification}

Model 1:

\begin{eqnarray}
&& \delta \bar{\rho} =  - 5 \times 10^{-3} \times \exp[-(r / 8 {\rm Kpc})^2]  \nonumber \\
&& S = 1, \nonumber \\
&& P = 0, \nonumber \\
&& Q = -0.6 \ln (1+r/ Kpc) \nonumber \\ 
&& \times \exp (-0.003 {\rm Kpc}^{-1} \times r).
\end{eqnarray}

Model 2:

\begin{eqnarray}
&& \delta \bar{\rho} =   1.14 \times 10^{-3} \times \exp[-(r / 9 {\rm Kpc})^2]  \nonumber \\
&& S = 1, \nonumber \\
&& P = 0, \nonumber \\
&& Q = -1.45 \ln (1+0.2 {\rm Kpc}^{-1} \times r) \nonumber \\
&&   \times \exp (-0.003  {\rm Kpc}^{-1} \times r).
\end{eqnarray}

\begin{figure}
\vspace{0.2cm}
\includegraphics[scale=0.5]{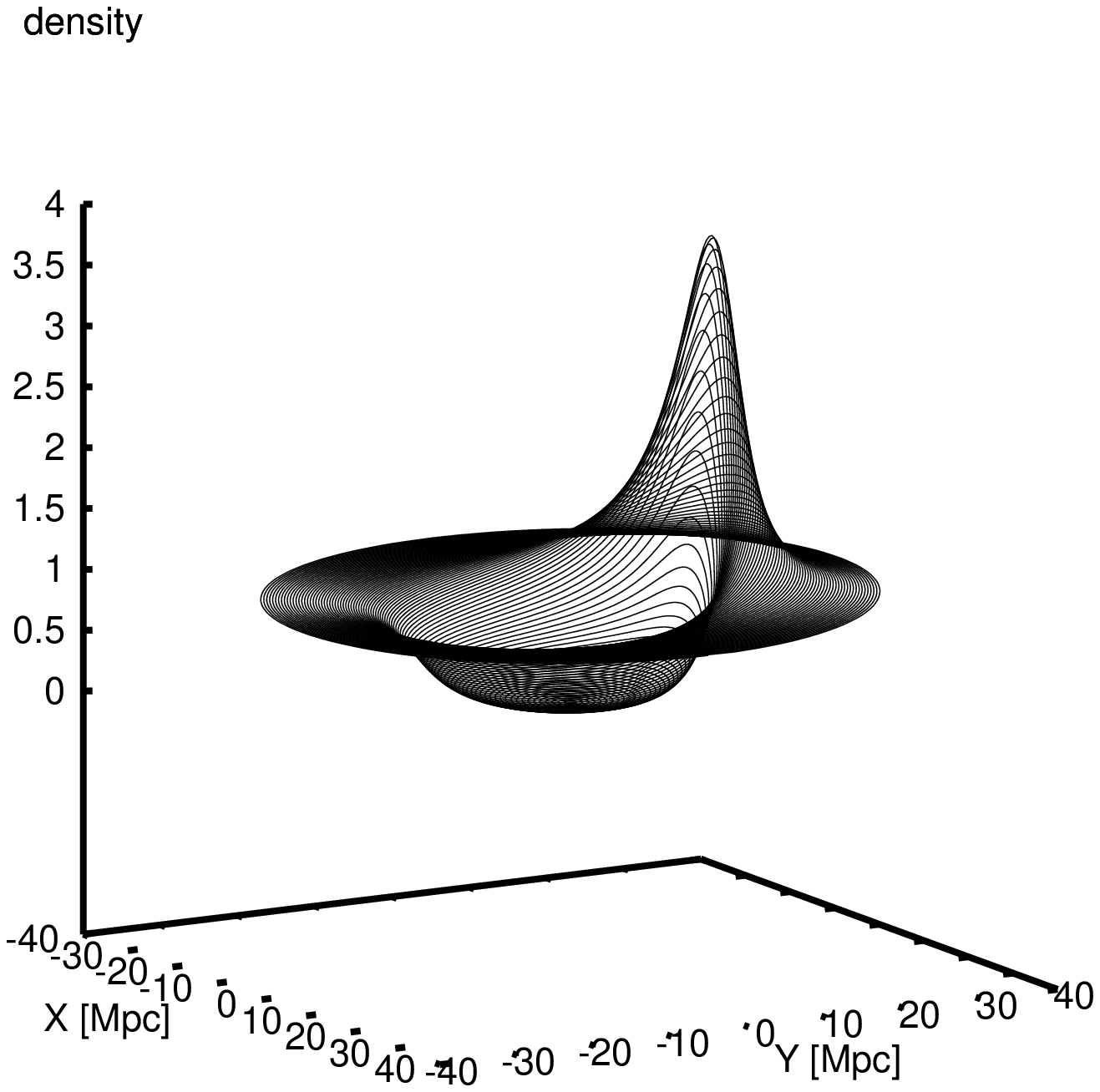}  
\includegraphics[scale=0.5]{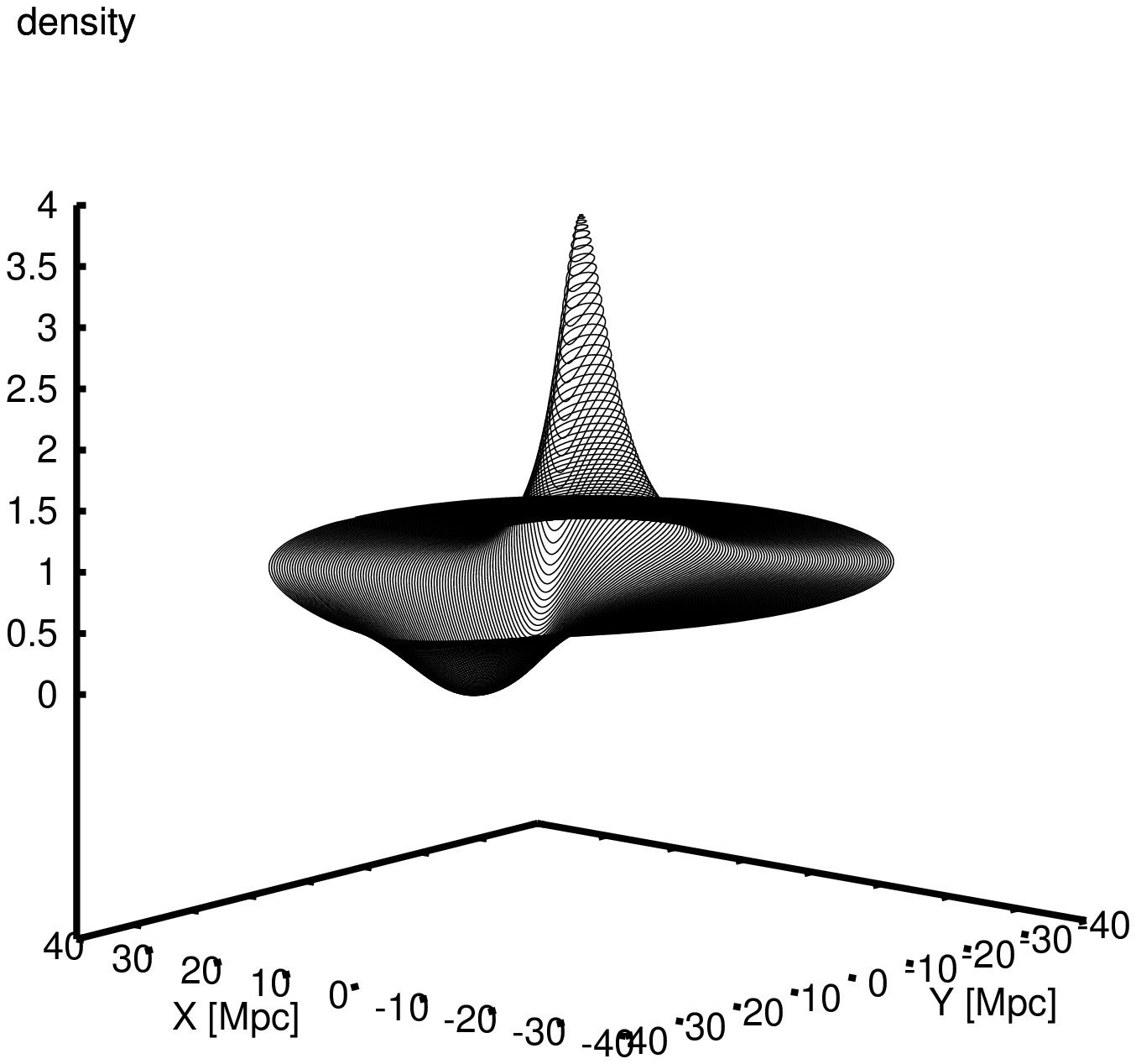}  
\caption{The present--day density distribution, $\rho/\rho_b$.
Upper panel presents model 1 ($\delta M<0$) and lower panel presents
model 2 ($\delta M>0$).}
\label{fig1}
\end{figure}

The density distributions of models
1 and 2 are presented in Fig. \ref{fig1}. As can be seen
the model with $\delta M <0$ has the void in the center, and the
supercluster is described by the dipole component of the density distribution.
In model 2 the converse applies.
The overdense region is at the origin and the void
is described by the dipole component of the density distribution.

\subsubsection{Evolution}\label{evoqnez}

In this section we compare the evolution of the density contrast, $\delta(t,r,\theta, \phi)$, and the 
$S_{2,1/2}(t, \Sigma)$ density indicator   for models 1,  2, with the corresponding models of a single void and with the models of a single supercluster
obtained within the Lema\^itre--Tolman model.
The Lema\^itre--Tolman model is considered because within this model
one can describe a single  spherically symmetric structure.
Such a comparison can demonstrate
how evolution of a structure
changes if there is another structure in its close proximity.

\begin{figure}
\includegraphics[scale=0.35, angle=270]{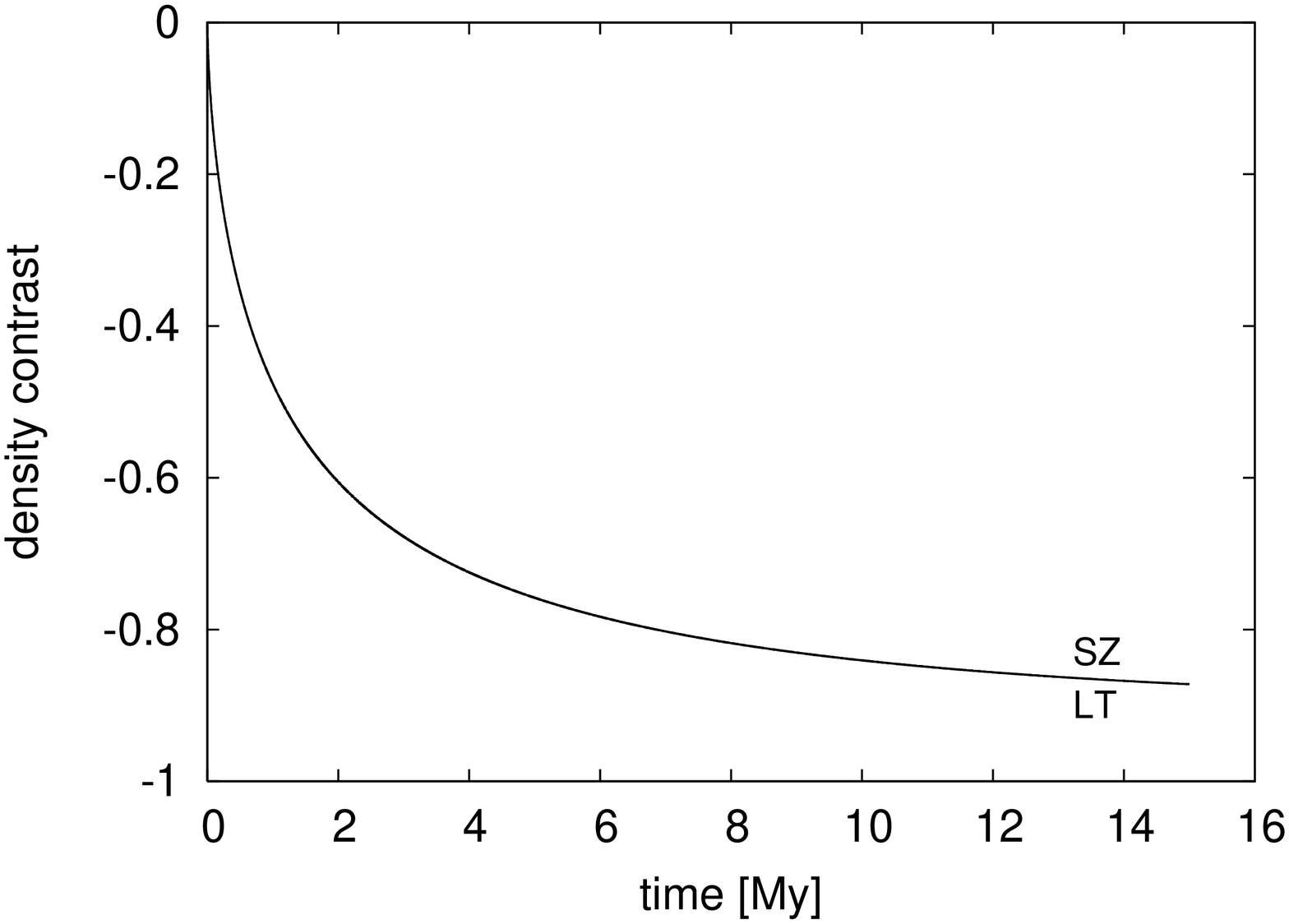}  
\includegraphics[scale=0.35, angle=270]{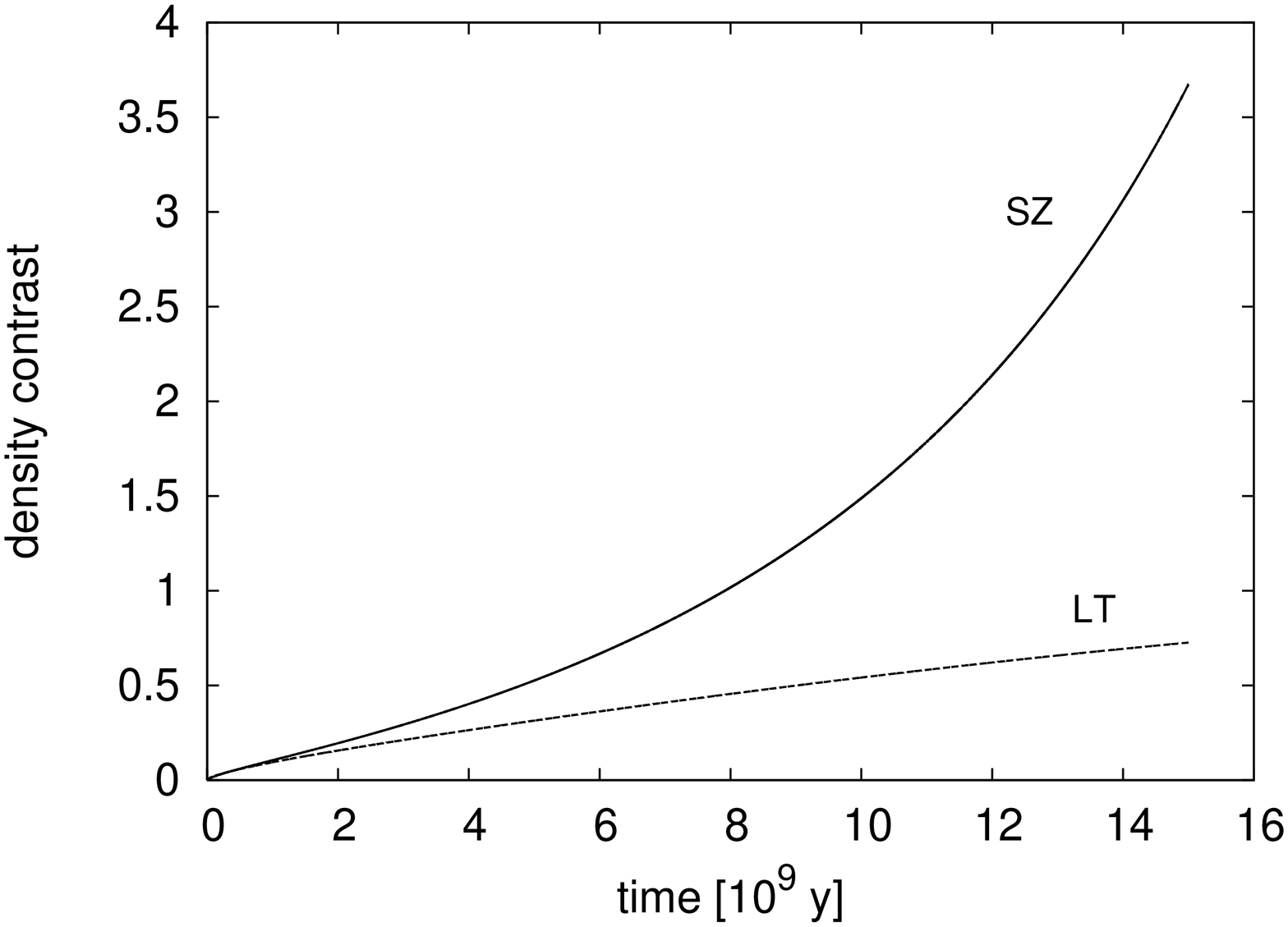}  
\caption{The evolution of the density contrast inside the void (upper panel), and inside the supercluster (lower panel) for model 1 ($\delta M<0$). The curve SZ presents the evolution 
within the Szekeres model; curve  LT presents the 
evolution within the Lema\^itre--Tolman model.}
\label{fig2}
\end{figure}

Fig. \ref{fig2} presents the evolution of the density contrast of the model 1 in comparison with corresponding 
models obtained within the Lema\^itre--Tolman model.
The Lema\^itre--Tolman model was specified by assuming the same condition 
as the ones in the Szekeres model at the initial instant.
Namely, the Lema\^itre--Tolman model was specified by 
$t_B = 0$ and the profile of the density distribution.
The local density contrast, $\delta$, is compared at the point of the 
maximal and minimal density value.
Upper panel of Fig. \ref{fig2} presents the evolution of the density contrast inside the void.
 As can be seen
the bahaviour of the density contrast in both models is similar.
This due to the regular conditions at the origin; ie.
at the origin, where $\Phi = 0$ and some other functions 
are also equal to zero, such conditions have to be imposed so that there would be no singularity at the origin (for a detailed description of the regularity conditions at origin see \cite{HK}).
These conditions imply that the origin behaves like a Friedmann model and this is the reason why the quasispherical 
Szekeres and Lema\^itre-Tolman models 
are of a very alike evolution pattern at the origin.
The lower panel of Fig. \ref{fig2}
presents the evolution of the density contrast 
at the very center of the overdense region
of the model 1 and the corresponding Lema\^itre--Tolman model. 
The growth of density contrast in the Szekeres model is much faster than in the corresponding 
Lemaitre--Tolman model.
The results of this comparison indicate that within the perturbed region of mass below the background 
mass ($\delta M <0$) the evolution of underdensities does not change but the evolution of the overdense regions situated at the edge of the  underdense regions is much faster than
the similar evolution of isolated structures.

\begin{figure}
\includegraphics[scale=0.35, angle=270]{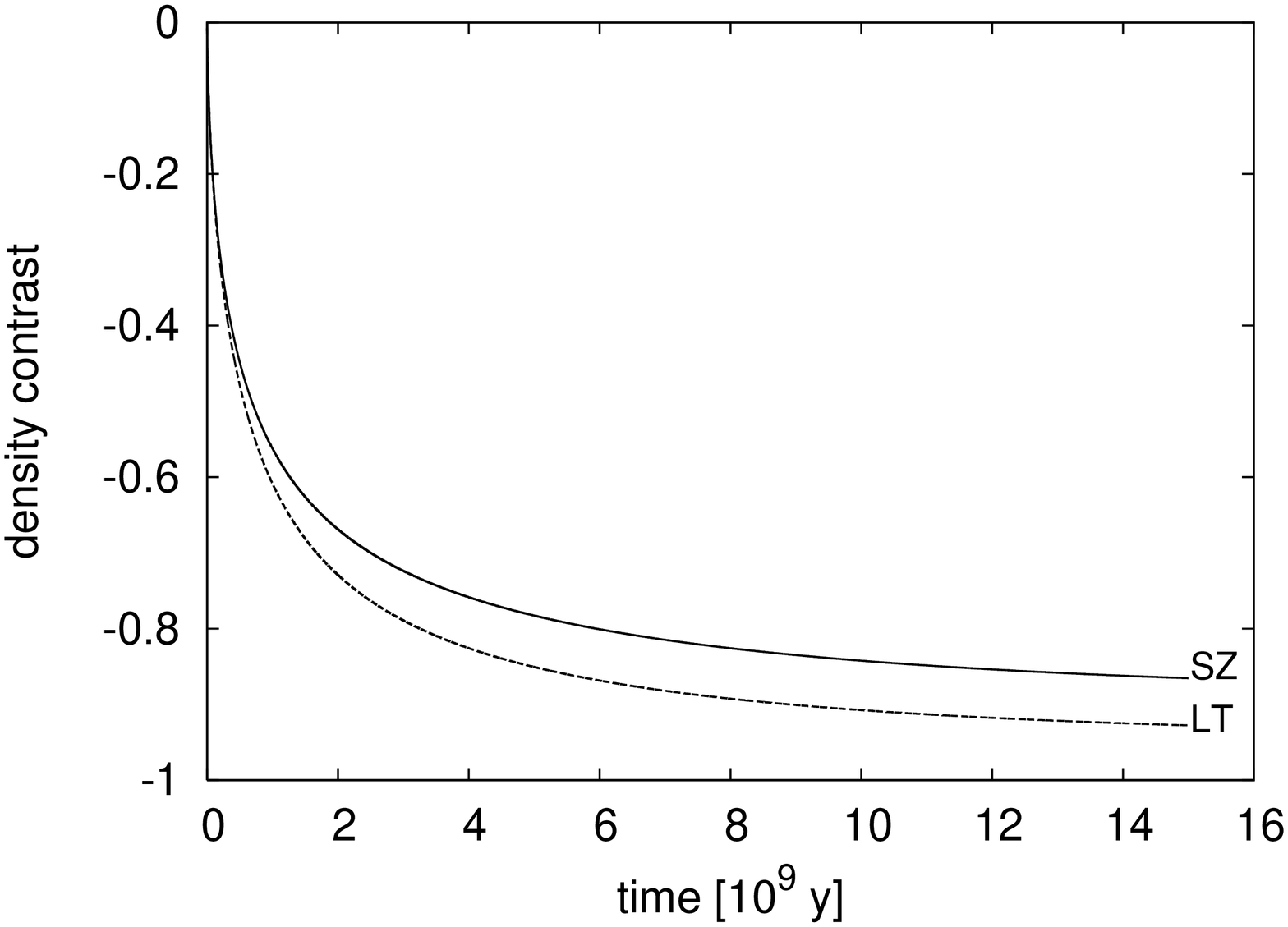}  
\includegraphics[scale=0.35, angle=270]{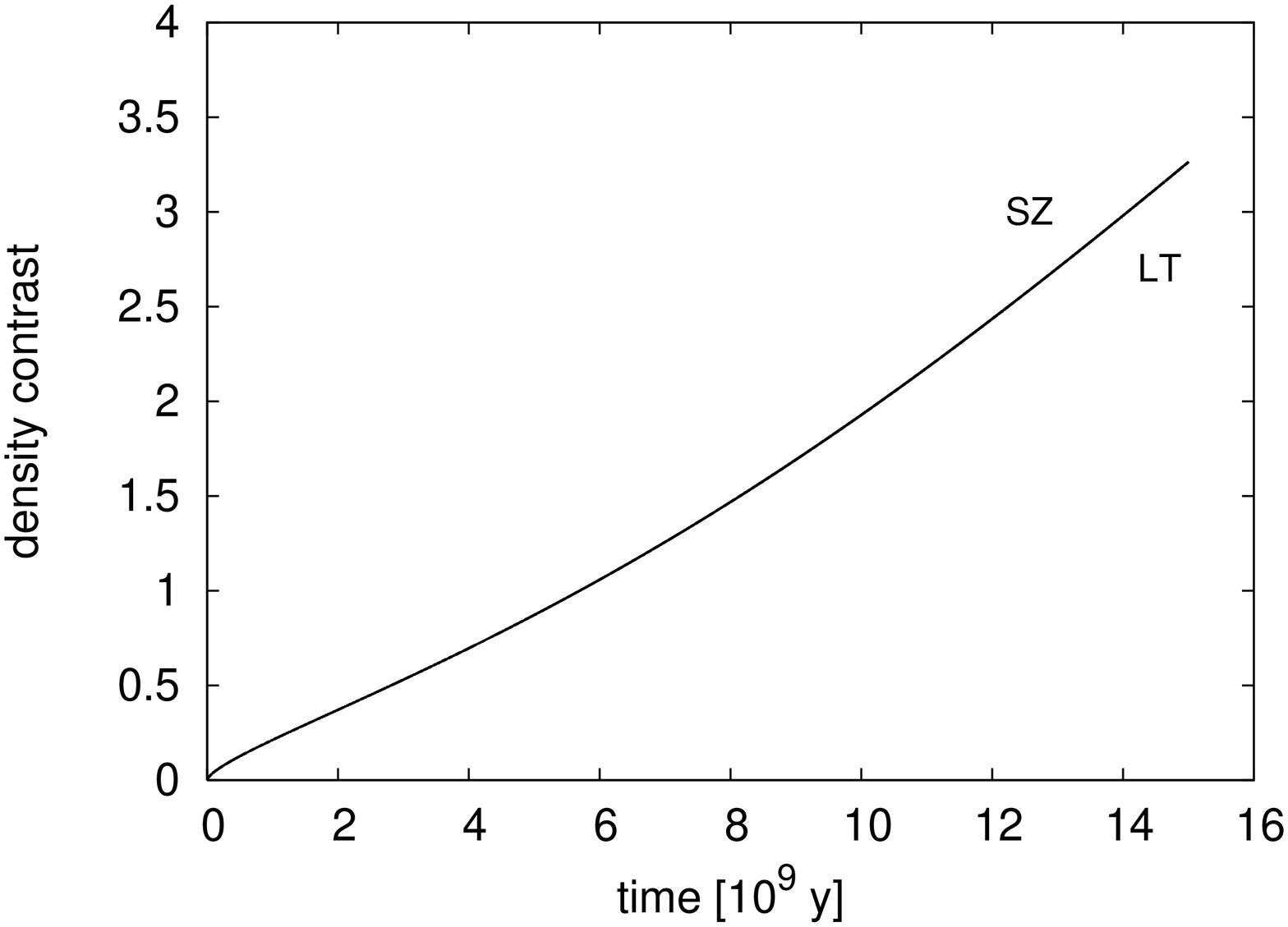}  
\caption{The evolution of the density contrast inside the void (upper panel), and inside the supercluster (lower panel) for model 2 ($\delta M>0$). The curve SZ presents the evolution 
within the Szekeres model; curve LT presents the 
evolution within the Lema\^itre--Tolman model.}
\label{fig3}
\end{figure}

The evolution of the density contrast of model 2 ($\delta M >0$) is presented in Fig. \ref{fig3}: the evolution of the density contrast at the point of minimal density is depicted in the upper
 panel of Fig. \ref{fig3}, and the evolution at the origin is depicted in the lower panel of Fig. \ref{fig3}.
Similarly as in model 1, the evolution at the origin in the Szekeres model and in the Lema\^itre--Tolman model are very alike.
The evolution of the void, however, is slower within the Szekeres model than it is in the Lema\^itre--Tolman model.
This implies that single, isolated voids evolve much faster than the ones which are in the neighborhood
of large overdensities where the mass of the perturbed region is above the background mass ($\delta M>0$).

Now, let us compare the evolution of the $S_{2,1/2}$ density indicator:

\begin{equation}
S_{2,1/2} = \int_{\Sigma} \sqrt{ \left| \frac{h^{\alpha \beta}}{\rho^2} \frac{\partial \rho}{ \partial x^{\alpha}}
\frac{\partial \rho}{ \partial x^{\beta}} \right|} \sqrt{-{\rm det} g} {\rm d} r {\rm d} \theta {\rm d} \phi,
\end{equation}

Similarly as above two different types of $\Sigma$ are going to be considered:

\[ \rho > \rho_b \rightarrow \Sigma = C, \]

\[ \rho < \rho_b \rightarrow \Sigma = V. \]

\begin{figure}
\includegraphics[scale=0.35, angle=270]{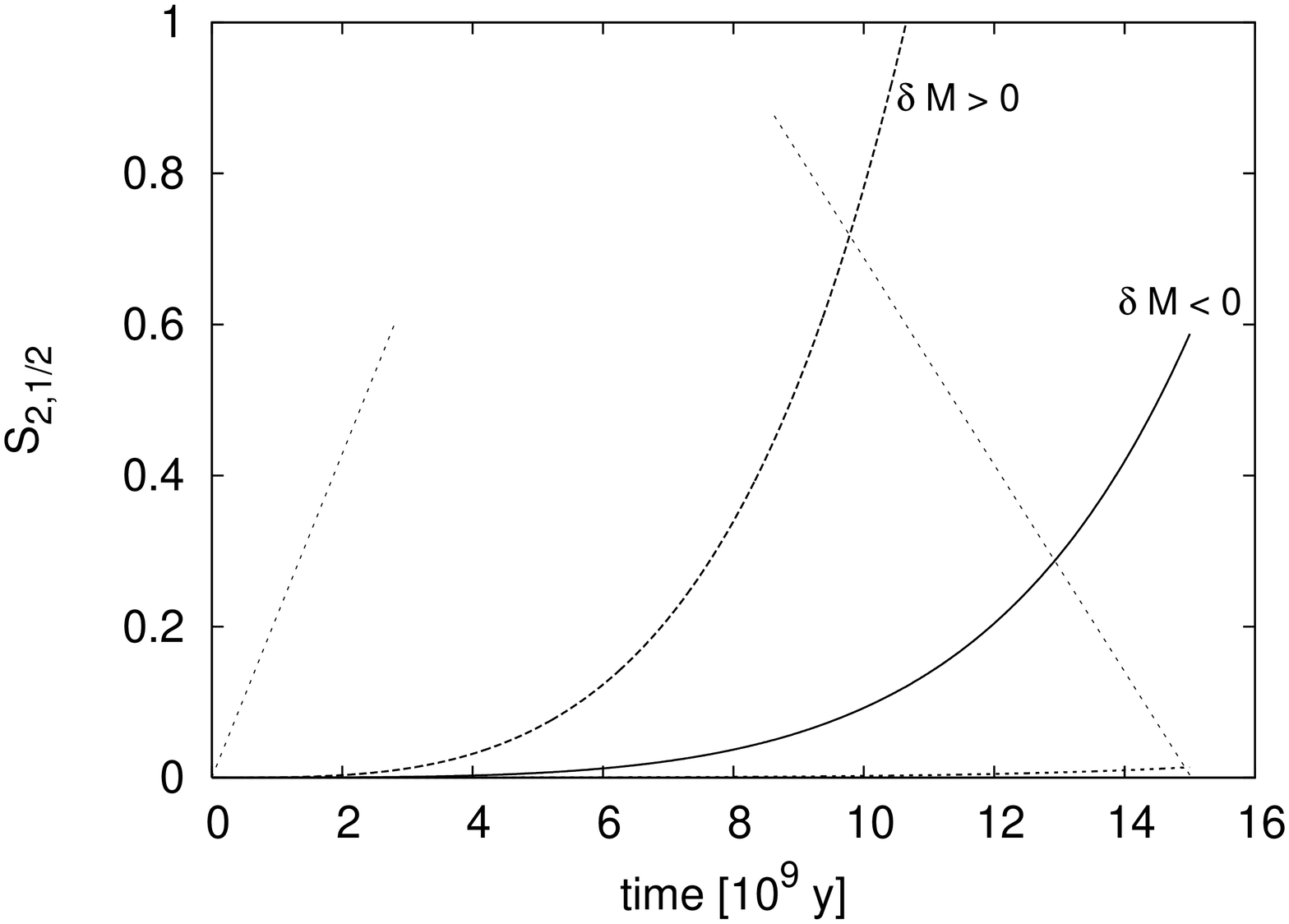}  
\includegraphics[scale=0.35, angle=270]{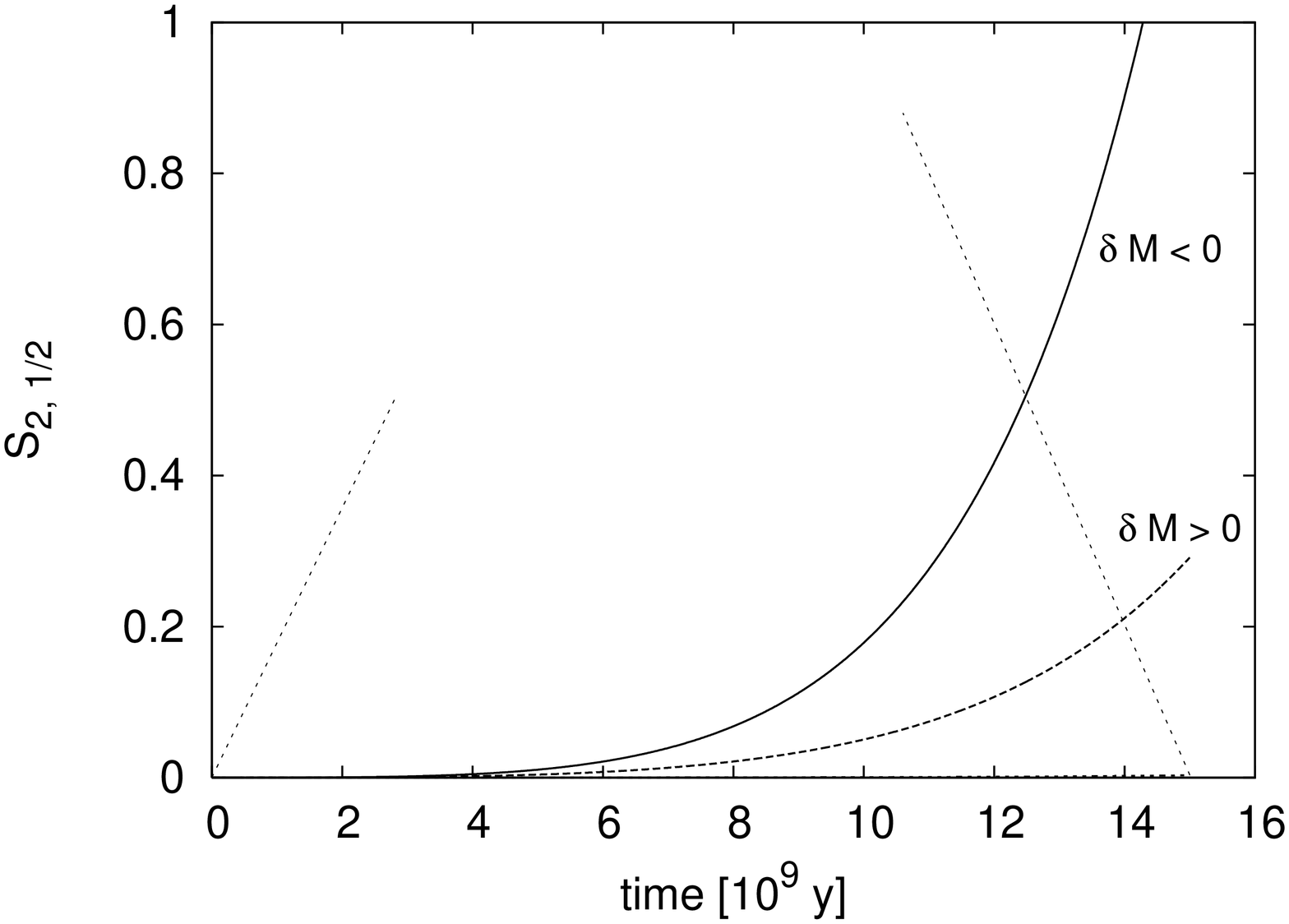}  
\caption{Comparison of $S_{2,1/2}$ for models with $\delta M>0$, $\delta M<0$, and corresponding LT model of a void in the upper panel and supercluster in the lower panel.}
\label{fig4}
\end{figure}

Since the value of $S_{IK}$ depends on units,
the results presented in Fig. \ref{fig4} 
were normalized so they are now of order of unity.

The upper panel of Fig. \ref{fig4} presents the evolution of $S_{2,1/2}$ for an underdense region. The lower panel of Fig. \ref{fig4} depicts the evolution of $S_{2,1/2}$ for an overdense region.
 As can be seen,  $S_{2,1/2}$ for the quasispherical Szekeres models considered  are comparable and 
the growth of $S_{2,1/2}$ for the Lema\^itre--Tolman model 
is much smaller.
This is because the volumes of the considered regions are diffrent. 
In the Szekeres model the volume is larger
than the volume in the Lema\^itre--Tolman model.

Figs. \ref{fig1}, and \ref{fig2} present the shape of the structures without corrections
for the shell displacement. For expample the void in Fig. \ref{fig1} and Fig. \ref{fig5} (upper panels) 
seems to be almost spherical. In fact this void is 
squeezed in the  
$+Y$ direction and elongated in the $-Y$ direction [$Q' \ne 0$, $P' = 0 = S'$ --- see the metric (\ref{ds2ss}) and Eq. (\ref{nuz})]. 
This fact also leads in some regions to 
density gradients larger than in the Lema\^itre--Tolman model,
hence, such a large disproportion in $S_{2,1/2}$ between the Szekeres model and 
the Lema\^itre--Tolman model.

 The results presented above indicate that
the evolution of the Szekeres model is much more complex than 
the evolution of the Lema\^itre--Tolman model. 
The evolution  not only depends on the value of the
density contrast but also on the density gradients and the volume of the perturbed region.This is the reason why $S_{2,1/2}$
of the void in model 2 ($\delta M >0$) is higher than in other models although the density contrast in this model evolved slower then in 
model 1. Similarly, as can be seen by comparison of  Figs \ref{fig5} and \ref{fig6},
the overdense region in the model with $\delta M < 0$ is much larger 
 than in other models, and as a consequence the $S_{2,1/2}$ for this   model evolves much faster than in other models.
The $S_{2,1/2}$ provide  us with information about the evolution of the whole perturbed region.

The evolution of  the density at single point 
is described with the local density contrast $\delta$.
 As can be seen
the evolution of the maximal and minimal density contrast 
depends on the value of $\delta M$ of perturbed region. 
The evolution of the density contrast inside large and isolated voids is faster than inside small voids which are surounded by highly dense regions. On the other hand, the evolution of the density contrast in highly  dense regions in close neighborhood of large 
voids is faster due to faster mass flow from the voids.

\subsection{Models with $P' \ne 0 \ne S',~Q' \ne 0$}\label{sqpnz}

In this section models of non--constant $P, Q$ and $S$  
are investigated.
The evolution of these models is compared with the evolution 
of models which were considered in Sec. \ref{p's'z}

\subsubsection{Models specification}

Model 3:

\begin{eqnarray}
&& \delta \bar{\rho} =  - 5 \times 10^{-3} \times \exp[-(r / 8 {\rm Kpc})^2]  \nonumber \\
&& S = -\left(r/{\rm Kpc} \right)^{0.4}, \nonumber \\
&& P = 0.55 \left(r/{\rm Kpc} \right)^{0.4}, \nonumber \\
&& Q = 0.33 \left(r/{\rm Kpc} \right)^{0.4}
\end{eqnarray}

Model 4:

\begin{eqnarray}
&& \delta \bar{\rho} =   1.14 \times 10^{-3} \times \exp[-(r / 9 Kpc)^2]  \nonumber \\
&& S = -\left(r/{\rm Kpc} \right)^{0.9}, \nonumber \\
&& P = 0.55 \left(r/{\rm Kpc} \right)^{0.8}, \nonumber \\
&& Q = 0.33 \left(r/{\rm Kpc} \right)^{0.8}
\end{eqnarray}

\begin{figure}
\includegraphics[scale=0.25]{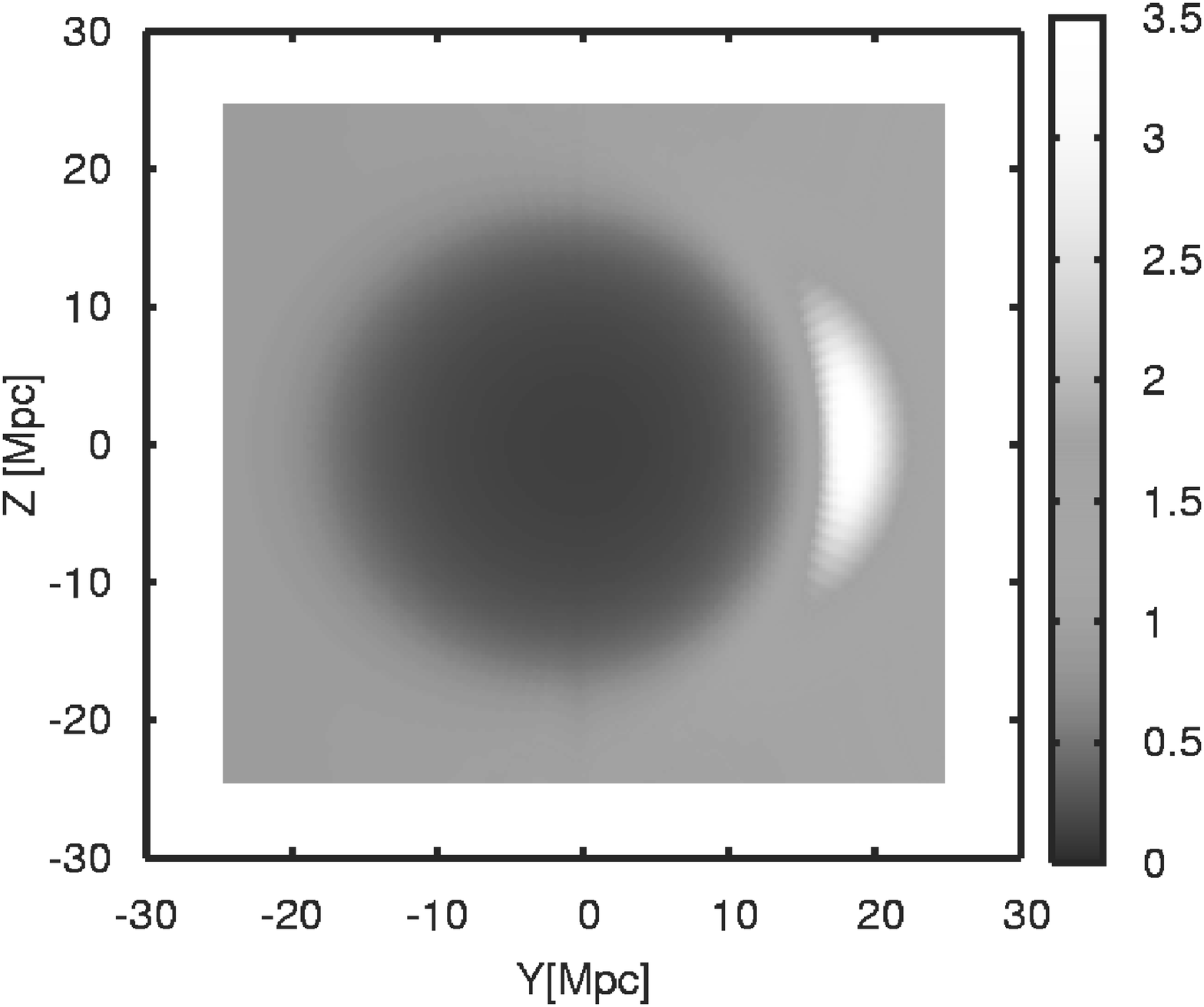}  
\includegraphics[scale=0.25]{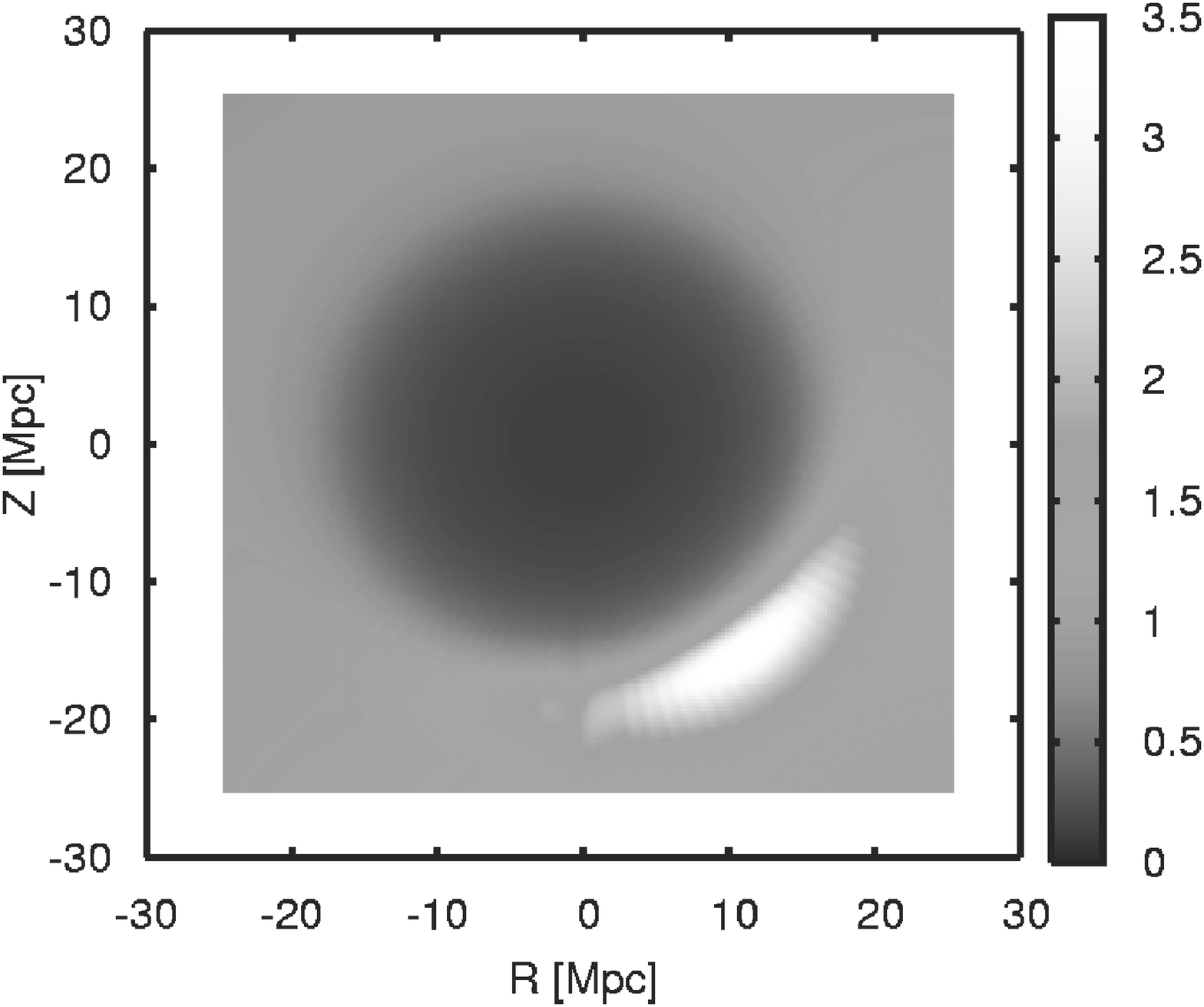}  
\caption{The present--day colour coded density distribution, $\rho/\rho_b$.
Models with $\delta M <0$. Upper panel --- $P' = S' = 0$.
Lower panel --- $P' \ne 0 \ne S'$.}
\label{fig5}
\end{figure}

\begin{figure}
\includegraphics[scale=0.25]{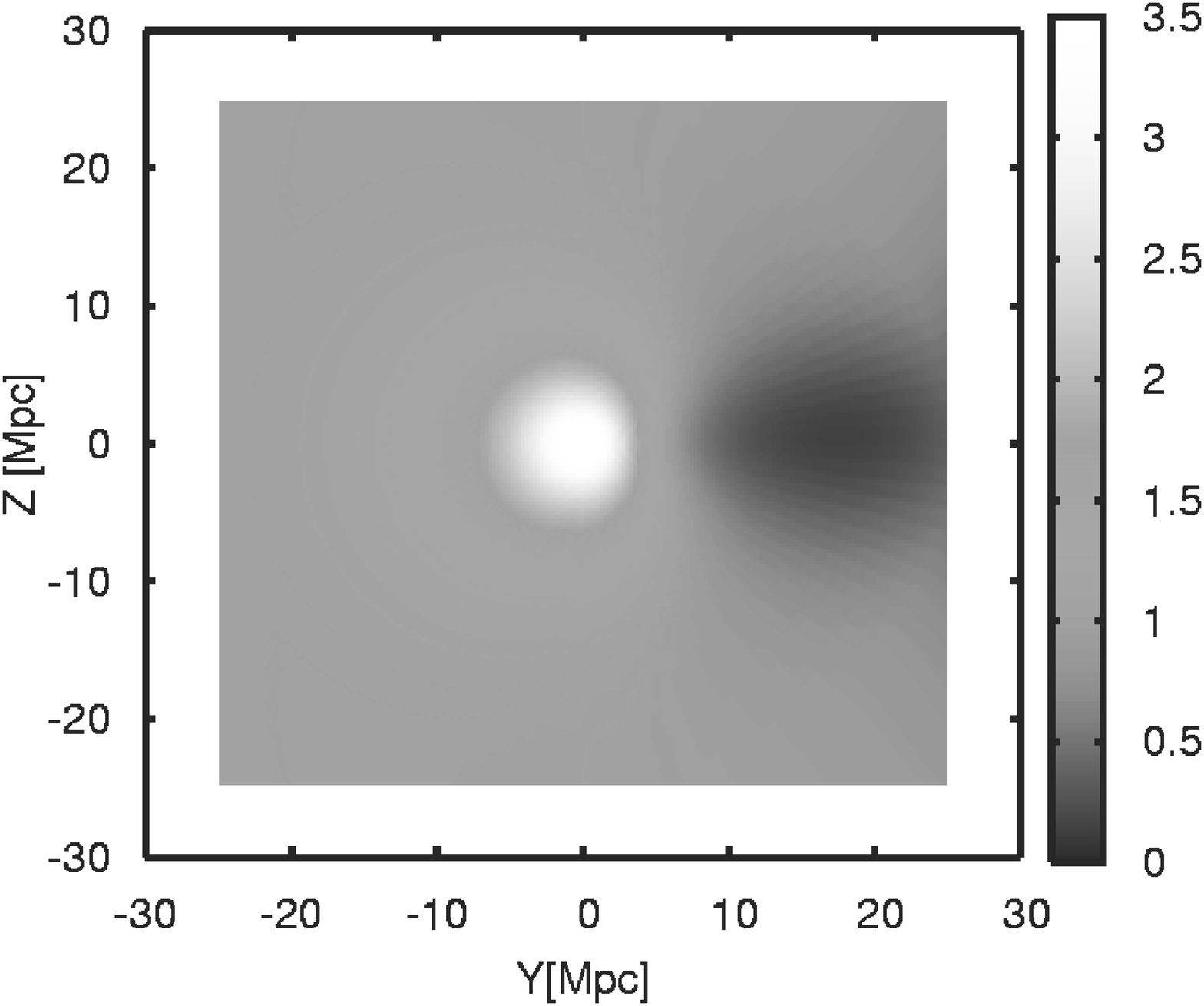}  
\includegraphics[scale=0.25]{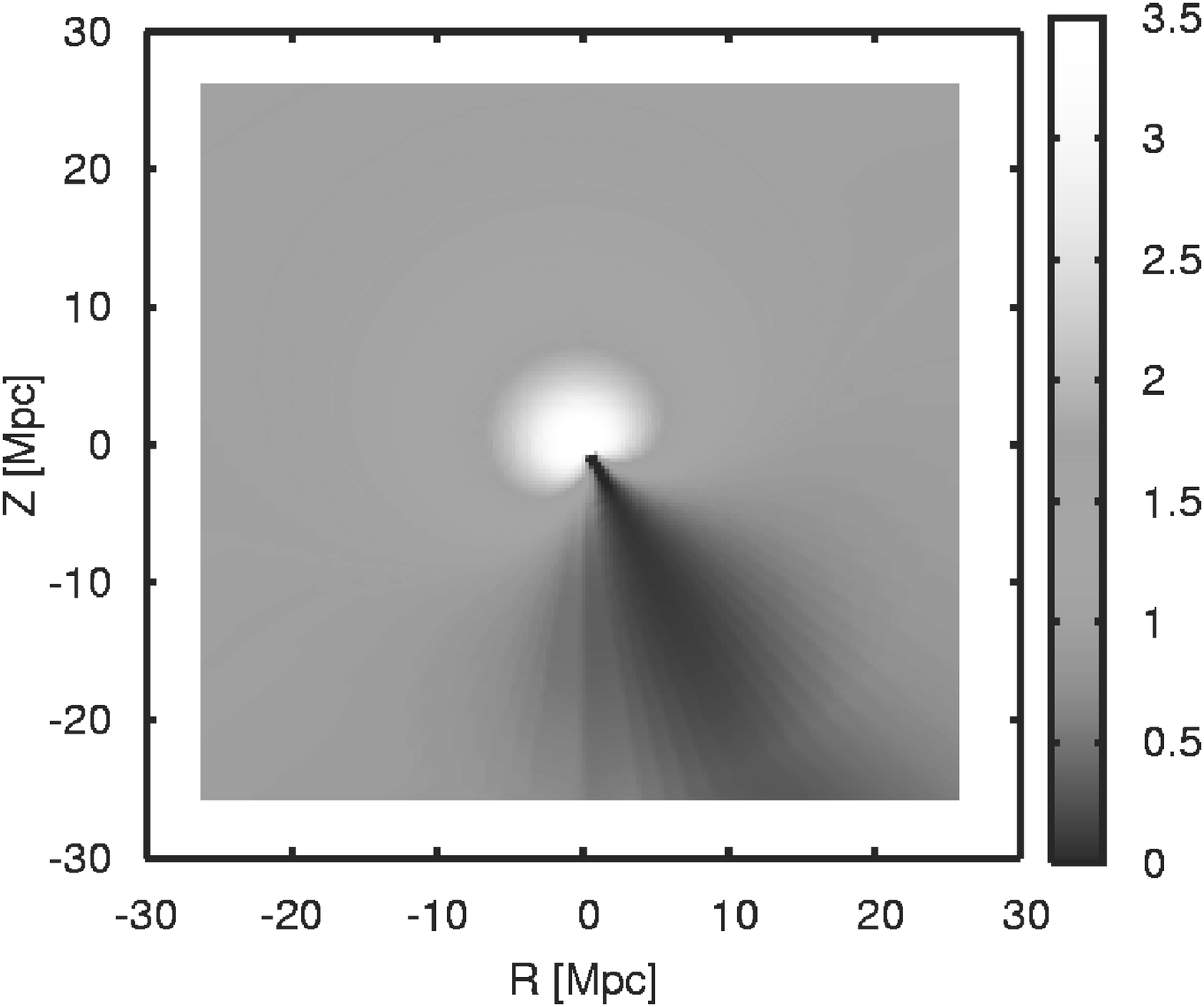}  
\caption{The present--day colour coded density distribution, $\rho/\rho_b$. Models with $\delta M > 0$. Upper panel --- $P' = S' = 0$.
Lower panel --- $P' \ne 0 \ne S'$.}
\label{fig6}
\end{figure}

Fig. \ref{fig5} presents the comparison of the present-day density distribution 
in models 1 and 3 in colour coded diagrams. It presents the vertical cross--sections of 
the considered
structures. The upper panel of Fig. \ref{fig5} presents the vertical cross--section
through the surface of $\phi = \pi/2$ and the lower panel presents
the cross section through the surface of $\phi \approx \pi/6$.
The comprehensive study of the 
 vertical and horizontal cross--sections
of similar models was presented by Bolejko \cite{kb3}.
Fig. \ref{fig6} also presents the vertical cross--sections 
of models 2 and 4.
As can be seen, both structures appear to be similar but,
in comparison with model 1, in model 3 the dipole component is moved down and right.
Model 4 on the other hand presents the structure moved down and right in comparison with model 2.

The next section discusses the evolution of these structures.

\subsubsection{Evolution}

The evolutions of the density contrast inside the voids and superclusters of models 3 and 1 is very similar which needn't be suprising as 
 model 3 has the same $\bar{\delta}(r)$ as model 1.
Also, the evolutions of the corresponding density contrasts of model 4 and 2
is similar. 
The functions $S,P,Q$ were chosen so that they reproduce the same shape of current
structures and the same density contrast inside them --- that
is why that the evolution of a local density contrast is comparable for models 1 and 3, and for models 2 and 4.
However, it is not clear whether or not the evolution of
$S_{2,1/2}$ is comparable too.
 When the functions $S, P, Q$ are not constant,
the axis of a density dipole changes. Also, the volume of the perturbed region 
as well as the density gradients can be different. So it may be interesting to compare the evolution of the whole perturbed underdense and overdense regions of models 1, 2, 3, and 4.

\begin{figure}
\includegraphics[scale=0.35, angle=270]{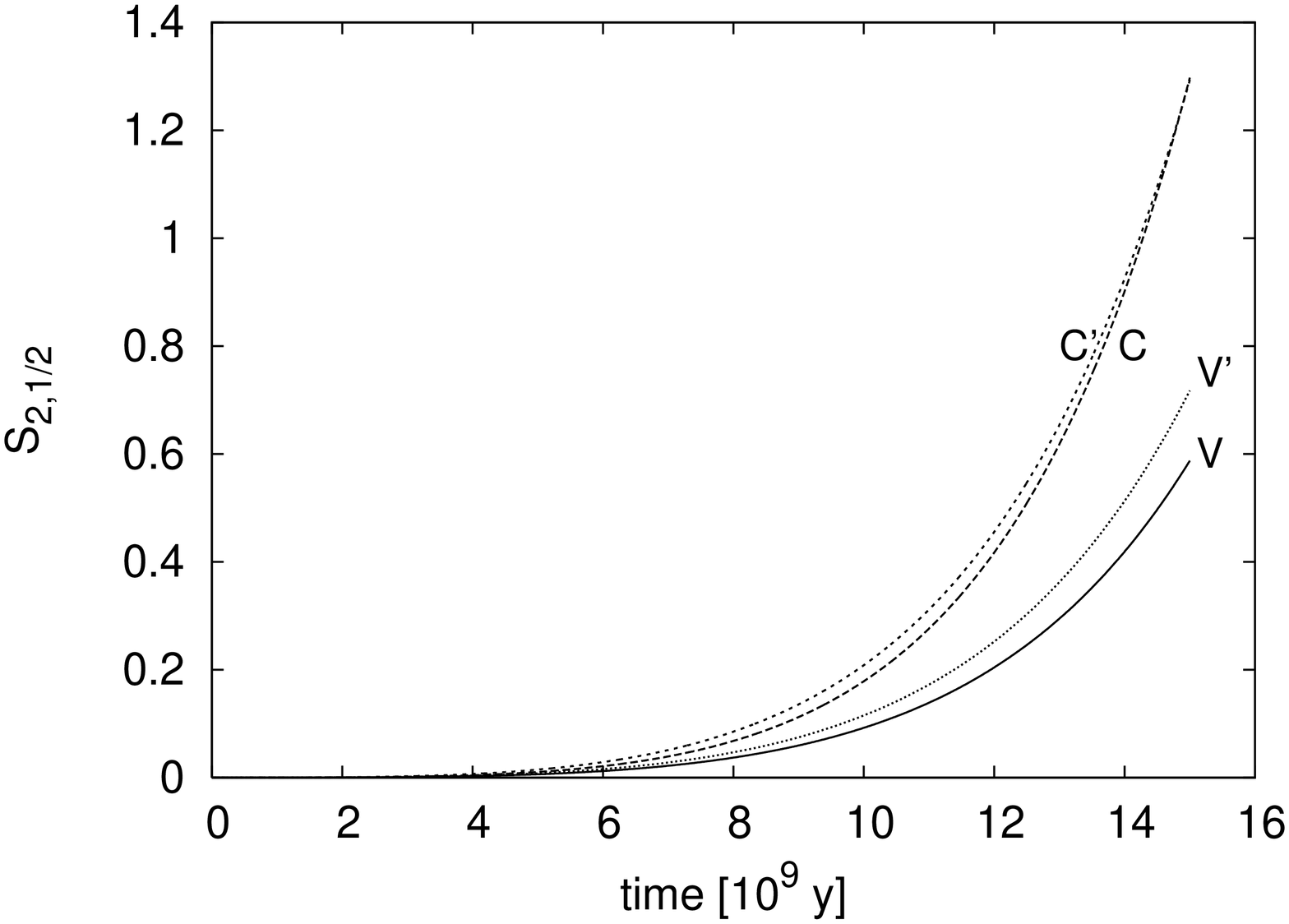}  
\includegraphics[scale=0.35, angle=270]{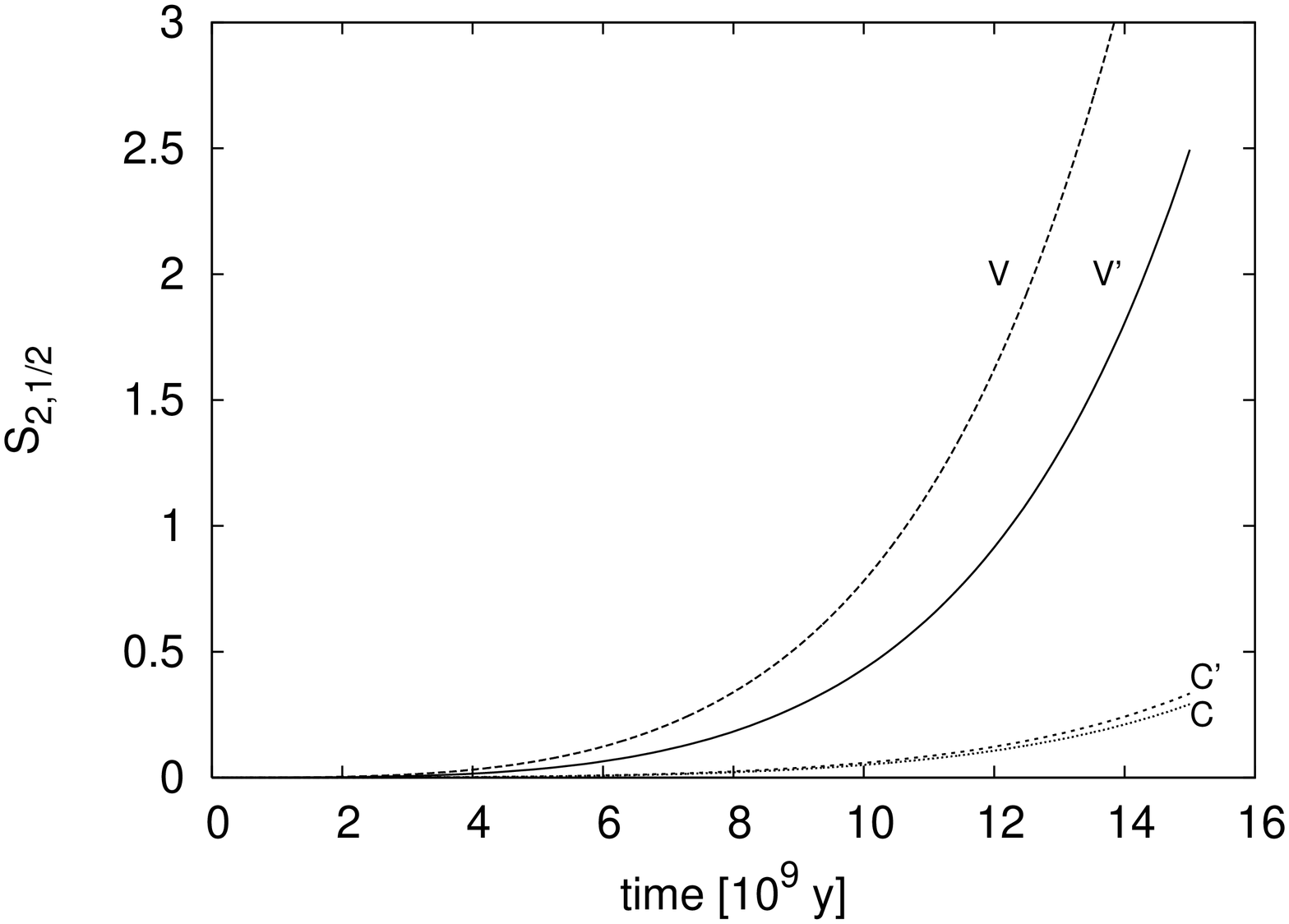}  
\caption{Comparison of $S_{2,1/2}$ for models with $\delta M<0$ (upper panel) and
with $\delta M<0$ (lower panel). C corresponds to a "supercluster" --- an overdense region, and V correspondes to a "void" --- underdense 
region. Primes denote models with $P' \ne 0 \ne S'$.}
\label{fig7}
\end{figure}

Fig. \ref{fig7} presents the comparison of $S_{2,1/2}$ evolution of model 1--4.
Similarly as in Fig. \ref{fig4},  the values of $S_{IK}$ 
were normalized so they are now of order of unity.
The primed letters denote models of $S' \ne 0 \ne P', Q' \ne 0$.
As can be seen the evolution of $S_{2,1/2}$ for all these models is also comparable.
These results imply that the evolution in the quasispherical Szekeres model does not depend on the position of the dipole component. As long as the shape
 and density contrast  of the analysed models are similar,
such models evolve in a very similar way.

\section{The role of expansion}\label{expan}

The faster or slower evolution rate of the previously presented models
is reflected by their current expansion rate. As has been shown above, 
the evolution does not depend on a relative position of the dipole
component (evolution of models 1 and 3 is similar).
 Thus, let us focus on model 1 and model 2 only.

\begin{figure}
\includegraphics[scale=0.29]{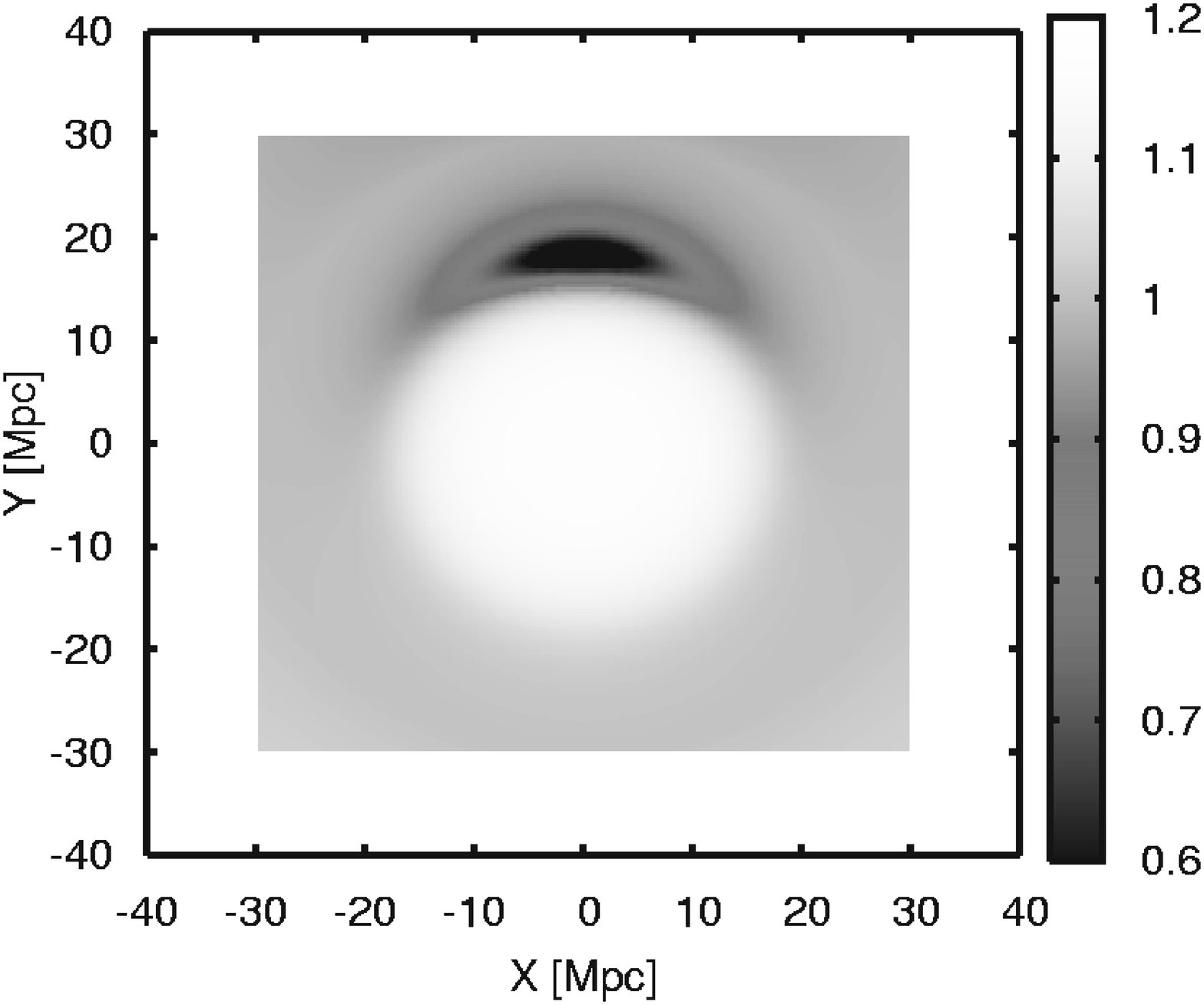}  
\includegraphics[scale=0.29]{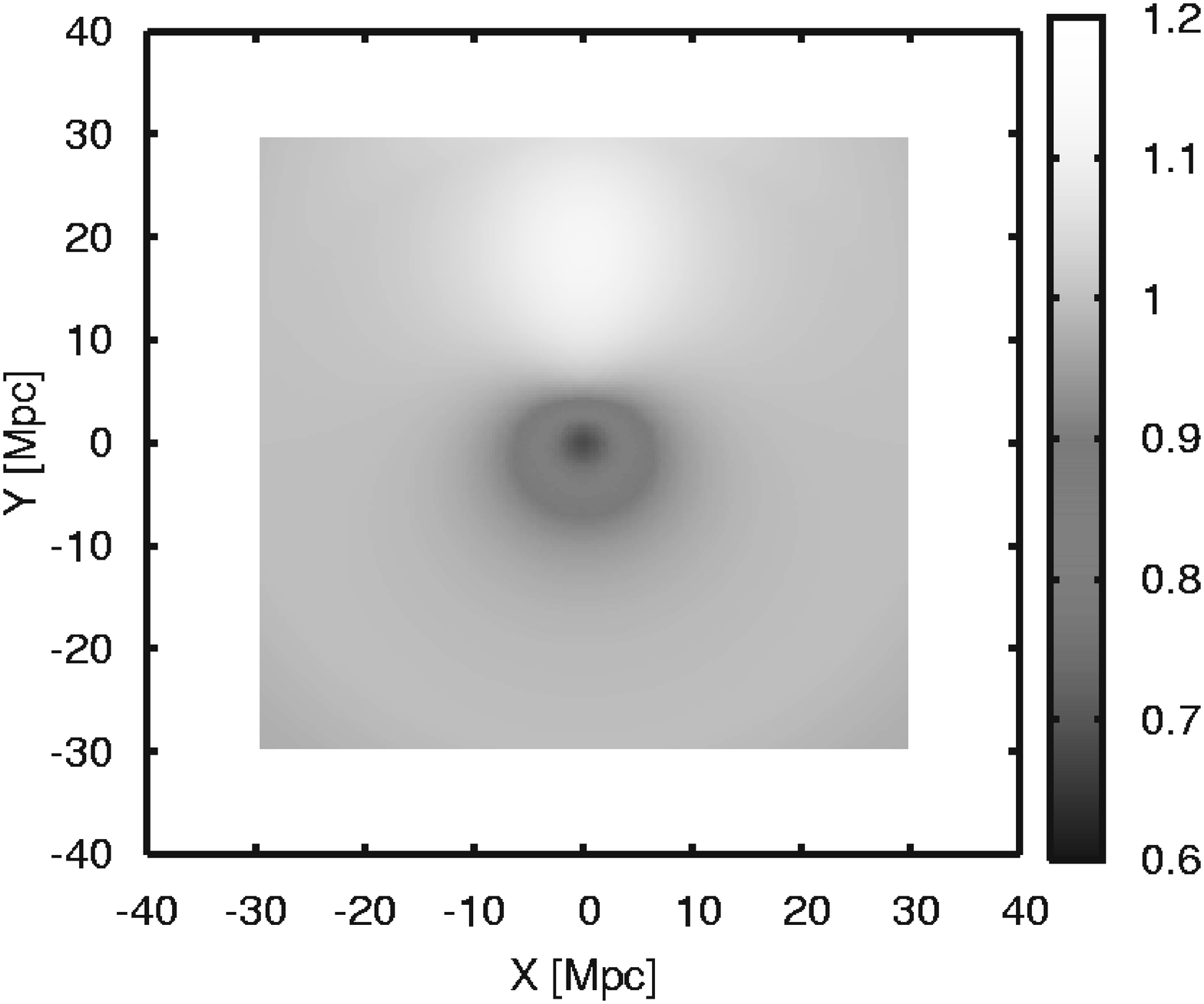}  
\caption{The $\Theta_{SZ} / \Theta_0$ ratio. Upper panel presents the
 ratio of model 1, lower panel presents the ratio of model 2.}
\label{fig8}
\end{figure}

Fig. \ref{fig8} presents the ratio, $\Theta_{SZ} / \Theta_0$, of the expansion parameter 
in the considered Szekeres models to the expansion parameter in 
the homogeneous background.
As can be seen, model 1, with $\delta M < 0$, has a larger amplitude of this ratio,
and the evolution of a supercluster in this model is much faster 
than in the corresponding Lema\^itre--Tolman model.
On the other hand, model 2 ($\delta M>0$) has smaller amplitude of the 
$\Theta_{SZ} / \Theta_0 $ ratio and within model 2 the evolution of the
 density contrast inside the  void was much slower
than in the Lema\^itre--Tolman model.
So clearly the rate of the evolution is connected with the rate of the expansion. This conclusion is also supported by the continuity equation [Eq. (\ref{coneq})].

Still, there remains a question whether the conclusions presented 
at the end of Sec. \ref{evoqnez} about the evolution of a density contrast being dependent on the mass of perturbed region,
are not limited to the class of models considered in this paper.
Are these conclusions general? How they are relevant to
 the real large-scale structure of the Universe?
These are the questions that are addressed 
in this and the next section of this paper.

The models presented above were defined by choosing
functions $t_B, M(r), S(r), Q(r), P(r)$.
As can be seen, the functions $S(r), Q(r), P(r)$ describe
the position of the dipole and even
with $S$ and $P(r)$ constant,  we are still able
to reconstruct the cosmic structures.
Moreover, the functions $S, P$, and $Q$ are chosen in such a way  
that they reproduce the present day cosmic structures.

The other functions which are chosen to specify the model
include  $M(r)$ and $t_B(r)$.
Is it also possible to choose other set of functions,
such as $k(r)$, and $t_B(r)$, or any other functions 
such as those described in Ref. \cite{KH2006}.

However, if we choose for example $k(r) = 0$ 
and the mass distribution as 
in model 1 or in model 2, we find
that the initial fluctuations diminish
with time. This is consistent with 
the bahaviour of the Lema\^itre--Tolman models,
where $t_B'$ is known to describe the decaying modes
and $k'$ --- growing modes \cite{Silk}.

Therefore, it is not the $\delta M$ 
but the $k$ function that is significant for the evolution.
This was also noticed in spherically symmetric models of structure formation.
In Ref. \cite{BKH} it was concluded
that the evolution of cosmic voids is generally driven by
the velocity fluctuations rather than by the density fluctuations.
In Ref. \cite{kb1}, where the function
$k$ could also change in time, the conclusion is similar, 
i.e.  it is this function that plays an essential role 
in the process of the structure formation.

However, we cannot take any arbitrary $k(r)$ 
and $M(r)$  (or any other function, instead of $\delta M$, 
defining the model), because such arbitrary pairs of functions
are in many cases 
"unnatural" and lead either to a large
amplitude of $t_B$  or to a shell crossing singularity
(one of the conditions to avoid the shell crossing is $t_B' <0$ --- see 
Ref. \cite{HK}),
and in most cases --- to both of these situations.
The large amplitude of $t_B$ is undesirable.
The function $t_B(r)$ describes the moment of initial singularity.
The  observations of the cosmic microwave background radiation (CMB)
indicate that the Universe was very homogeneous at the last scattering moment and 
as a consequence the bang time function, $t_B(r)$, cannot 
have a lager amplitude than a few thousand years.
Larger values of $t_B(r)$ would induce temperature fluctuations on the CMB sky larger than observed.
On the other hand, models with $t_B =$ const are known
to describe growing modes only \cite{Silk},
so the assumption that $t_B =$ const seems very natural.
If we set $t_B$ and specify $M(r)$, then $k(r)$ is
already specified by the Einstein equations. On the contrary,
if we set $t_B$ and specify $k(r)$ then the $M(r)$
is already fixed.
Therefore,  for the class of models which evolve from small initial
fluctuations and do not have shell crossings
during the evolution, as well as reproduce structures
similar to the observed cosmic structures, 
the conclusions drawn at the end of Section \ref{evoqnez} are valid.
However,  it can now be seen
that it is not the mass fluctuations that matter
but the expansion rate.
Higher mass in the perturbed region 
slows down the expansion rate --- 
this is a condition hindering the evolution of cosmic voids.
On the other hand, if the mass of perturbed
region is below the background mass, such region expands
much faster than the background, leading 
to the formation of the large underdense regions.
Such large voids enhance the formation
of large elongated overdensities
formed at the edges of voids, 
which are usually called walls.

\section{Connection to the large-scale structure of the Universe}\label{tripstr}

The models presented above are models 
of two structures embedded in the homogenueous universe.
Although the Universe is much more complicated
than that, such simple models enable us to come to some
general conclusions. In this section the triple structure
is considered and it will be seen that the
behaviour of the evolution of cosmic structures
in this model is similar to this observed in the previous models.

Model 5 is specified by the following set of functions:

\begin{eqnarray}
&& t_B = 0, \nonumber \\
&& \bar{\delta} =  1 \times 10^{-3} \times \exp[-(r / 20 {\rm Kpc})^2] 
 - 6.5 \times 10^{-4}   \nonumber \\
&& \times \exp\{-[(r-35{\rm Kpc})/ 10 {\rm Kpc})^2]\}, \nonumber \\
&& S = 1, \nonumber \\
&& P = 0, \nonumber \\
&& Q = 0.33 \left(r/{\rm Kpc} \right)^{0.8}
\end{eqnarray}

\begin{figure}
\hspace{0.6cm}
\includegraphics[scale=0.24]{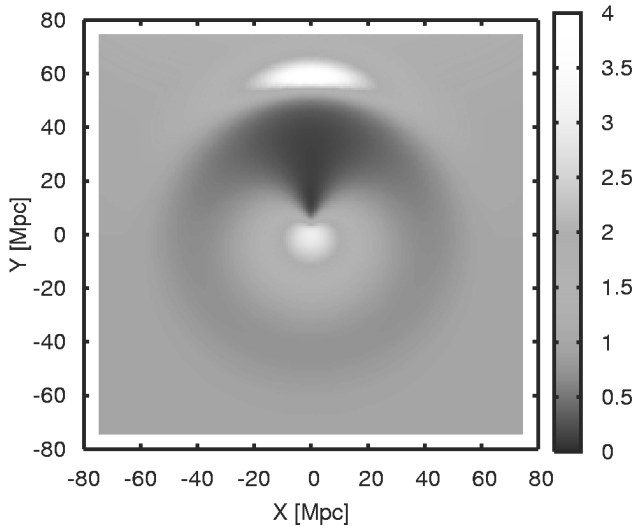}  
\caption{The present--day colour coded density distribution, $\rho/\rho_b$,
of model 5.}
\label{fig9}
\end{figure}

Fig. \ref{fig9} presents the density distribution. There is an
overdense region at the origin,
followed by a small void which spreads to a larger r.
At a larger distace from the origin, the void is  huge
and its larger side is adjacent to an overdense region.

This is diametrically different to what happened
close to the origin where the void adjourns to 
the  supercluster only with a narrow cusp.

\begin{figure}
\hspace{0.6cm}
\includegraphics[scale=0.35, angle=270]{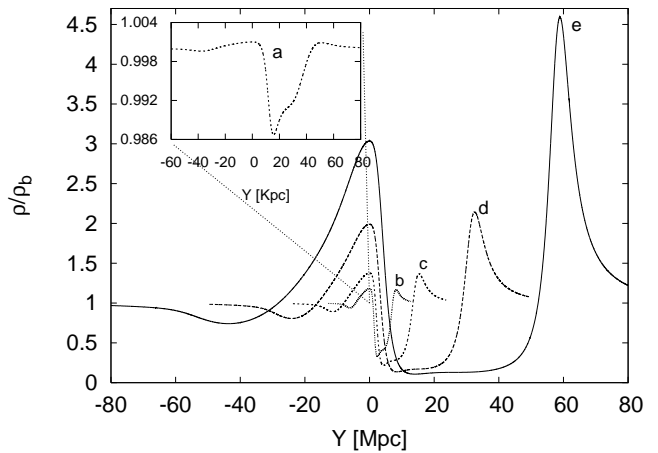}  
\caption{The evolution of a density profile of model 5. The profiles
correspond to the profile of $X = 0$ in Fig. \ref{fig9}.
Letters correspond to different time instants: a --- $0.5 \times 10^6$ after
the big bang; b --- $1.5 \times 10^9$yr; c --- $5 \times 10^9$yr; d --- $10 \times 10^9$yr; e --- present instant.}
\label{fig10}
\end{figure}

The evolution of model 5 is presented in Fig. \ref{fig10}. For clarity
Fig. \ref{fig10} presents only the profile which is represented
by the $X=0$ line in  Fig. \ref{fig9}. This profile 
is shown  for 5 diffrent time instants. 
As can be seen,  at a larger distance from the origin, where the void is large,
it   evolves much faster and exceeds the speed of the evolution
of the underdense region close to the origin. 
Another significant fact is that the overdense region 
connected by  the  void across a larger area
 evolves much faster
than the supercluster at the origin which is more compact.
This model exhibits the features of the models previously considered.
Thus, it might be speculated that the evolution
of the real structures follows similar patterns. Namely,
small voids in the Universe which are surrounded by 
large high density regions evolve much slower than the large isolated
voids. From the perspective of the continuity equation the expansion of the space
in this region is very slow and this is the reason why the voids do not
evolve as fast as they could. 
On the other hand, the expansion is much faster inside large voids,
where the mass of the perturbed region is below the background mass ($\delta M<0$). 
In these situations matter flows from central parts of the voids towards the highly dense regions which form at the voids' larger sides and enhance their evolutions.
In these situations matter flows from central parts of the voids towards the highly dense regions which form at the voids' larger sides and speed up their evolutions.

\section{Conclusions}\label{concl}

The   galaxy redshift surveys show that  the Universe is patchy with various structures.
These structures include small voids among compact clusters, superclusters and large voids surrounded by large walls or long filaments.

The evolution of these cosmic structures 
in different environments in the quasispherical Szekeres model was investigated.
The Szekeres model is one of the most complex and
spatially inhomogeneous exact solutions 
of the Einstein field equations and it has a potential to be
more widely used in cosmology.
Since it is an exact
solution of Einstein's equations, it 
enables to investigate the evolution of cosmic 
structures without such 
 approximations as linearity and small 
value of density contrast.
Moreover, the Szekeres model is flexible enough to 
describe more than one structure.

Having investigated various models with two 
or three structures within one frame 
it may be concluded that 
the evolution of the cosmic structures 
depends on the environment.
In perturbed region which mass is below the
background mass the amplitude of the
expansion's fluctuations is large
and as can be seen from 
the continuity equation [Eq. (\ref{coneq})],
such conditions enhance the evolution of cosmic structures.

The analyses presented in this paper indicate that small voids 
among large overdense regions
do not evolve as fast as the large voids do.
This is because the expansion of the space is faster inside large voids than inside smaller voids.
Moreover, this higher expansion rate 
inside the large voids leads to the formation
of large and elongated structures such as walls and filaments
which emerge at the edges of these large voids.

\section*{Acknowledgments}
I would like to thank Andrzej Krasi\'nski and Charles Hellaby for their valuable comments and discusions concerning the Szekeres model. Andrzej Krasi\'nski 
and Paulina Wojciechowska are 
 gratefully acknowledged for their help with preparing the manuscript.
This research has been partly supported by Polish Ministry of Science and Higher Education under grant N203 018 31/2873, allocated for the period 2006-2009.

\end{document}